%% file: main.tex
\documentclass[11pt,english]{article}
\pdfoutput=1
\usepackage[T1]{fontenc}
\usepackage{amssymb}
\usepackage{amsmath}
\usepackage[unicode=true,
bookmarks=true,bookmarksnumbered=false,bookmarksopen=false,
 breaklinks=false,pdfborder={0 0 1},backref=false,colorlinks=true]
 {hyperref}
\usepackage[hang,flushmargin]{footmisc}
\usepackage{relsize}
\usepackage{color}

\usepackage{amsfonts}
\usepackage{graphicx}
\usepackage{longtable}

\usepackage{lineno}
\usepackage{tcolorbox}
\tcbuselibrary{breakable}
\usepackage{placeins}

\DeclareFontShape{OT1}{cmr}{mx}{n}%
{<->cmr10}{}

\usepackage{tocloft}
\usepackage{etoolbox}

\hypersetup{
    colorlinks=true,
    linkcolor=blue,  
    urlcolor=blue,
    citecolor=blue
}

\makeatletter
\pretocmd{\l@section}{\hypersetup{linkcolor=black}}{}{}
\apptocmd{\l@section}{\hypersetup{linkcolor=blue}}{}{}
\pretocmd{\l@subsection}{\hypersetup{linkcolor=black}}{}{}
\apptocmd{\l@subsection}{\hypersetup{linkcolor=blue}}{}{}
\makeatother

\usepackage{amsmath,amssymb,amsthm,mathtools}
\numberwithin{equation}{section}
\usepackage{tikz-cd}
\let\sl\relax

\usepackage[normalem]{ulem}
\definecolor{JBDComment}{rgb}{0.50,0.20,0.65}

\definecolor{CCComment}{rgb}{0.00,0.50,0.30}

\definecolor{JPMComment}{rgb}{0.85,0.33,0.10}

\definecolor{RTComment}{rgb}{0.86,0.10,0.10}

\newcommand{\AuthorEdits}[1]{%
  \expandafter\newcommand\csname#1replace\endcsname[2]{\textcolor{#1Comment}{\ifstrempty{##1}{##2}{\sout{##1}\ifstrempty{##2}{}{ ##2}}}}%
}
\forcsvlist{\AuthorEdits}{JBD,CC,JPM,RT}

\newcommand{\C}[0]{\mathcal C}

\newcommand{\zee}[1]{\mathbb Z_{#1}}

\newcommand{\sl}[1]{\mathrm{SL}(2,#1)}
\newcommand{\mods}[0]{\mathfrak s}
\newcommand{\modt}[0]{\mathfrak t}
\newcommand{\irr}[0]{\text{Irr}}
\newcommand{\Uone}{\text{U}(1)}

\newcommand{\SU}{\text{SU}}
\newcommand{\Spin}{\text{Spin}}

\newcommand{\bra}[1]{\left\langle #1 \right\rvert}
\newcommand{\ket}[1]{\left\lvert #1 \right\rangle}

\usepackage[
  backend=bibtex,
  style=phys,
  biblabel=brackets,
  sorting=none,
  eprint=true,
  doi=true,
  url=false,
  isbn=false,
  giveninits=true,
  maxbibnames=10,
]{biblatex}
\usepackage{subcaption}
\usepackage{cleveref}
\crefname{figure}{Fig.}{Figs.}
\Crefname{figure}{Fig.}{Figs.}
\crefname{equation}{Eq.}{Eqs.}
\Crefname{equation}{Eq.}{Eqs.}
\crefname{section}{Section}{Sections}
\Crefname{section}{Section}{Sections}
\crefname{subsection}{Section}{Sections}
\Crefname{subsection}{Section}{Sections}
\crefname{appendix}{Appendix}{Appendices}
\Crefname{appendix}{Appendix}{Appendices}
\newtheorem{conjecture}{Conjecture}
\crefname{conjecture}{Conjecture}{Conjectures}
\AddToHook{cmd/appendix/after}{%
  \crefalias{subsection}{appendix}%
  \crefalias{subsubsection}{appendix}%
}
\AtBeginDocument{}

\input{figures/tikz-diagrams/ribbon-twist-diagrams.tex}

\input{figures/tikz-diagrams/s-matrix-diagrams.tex}

\begin{document}
\hypersetup{citecolor=blue,linkcolor=blue,urlcolor=blue}
\interfootnotelinepenalty=10000
\baselineskip=18pt

\vspace*{2cm}
\thispagestyle{empty}

\begin{center}
{\fontsize{20.74pt}{0pt}  \bf
Learning the Anyon Atlas
}


\bigskip\vspace{1cm}{
{\large Jonah Berean-Dutcher$^{\,1,\,2}$, Clay C\'ordova$^{\,3}$, Julio Parra-Martinez$^{\,2}$,\\ Richard Tuddenham$^{\,3}$, and Zekai Zhu$^{\,4,\, 5}$}
} \\[7mm]
{$^{1}$\textit{University of British Columbia, Vancouver, V6T 1Z1, Canada} \\
$^{2}$\textit{Institut des Hautes \'Etudes Scientifiques, 91440 Bures-sur-Yvette, France} \\
$^{3}$\textit{Leinweber Institute for Theoretical Physics  \& Enrico Fermi Institute, University of Chicago} \\
$^{4}$\textit{School of Computer Science and Technology, Tongji University} \\
$^{5}$\textit{Project Numina}} \\
\end{center}

\bigskip
\centerline{\large\bf Abstract}
\begin{quote} \small
We extend our previous search for the modular data of topological quantum field theories in 2+1 dimensions~\cite{Berean-Dutcher:2026fck} from 22 to 42 anyons, identifying 150,679 new modular data. We do so by adding to our algorithm three rejection gates, two of them based on logistic regression, that discard representations of $\sl{\zee{}}$ unlikely to yield modular data. We estimate that roughly $80\%$ of the modular data between ranks 23 and 42 has been identified after closing under known constructions. With this data, we find counterexamples to a conjecture of Geiko and Moore that the reality of the higher Gauss sums suffices for time-reversal symmetry in Chern--Simons theories, and propose a refined criterion. 
We demonstrate the scalability of this gated implementation by identifying prime modular data at rank 100 that is not realized by any Chern--Simons theory.
The identified data through rank 42 has been made available at \href{https://AnyonAtlas.org}{\texttt{AnyonAtlas.org}}, to which further data will be added as it is found.
\end{quote}
\setcounter{footnote}{0}

\vfill
{\small \textit{E-mail:} \href{mailto:jbd@phas.ubc.ca}{\texttt{jbd@phas.ubc.ca}}, \href{mailto:clayc@uchicago.edu}{\texttt{clayc@uchicago.edu}}, \href{mailto:julio@ihes.fr}{\texttt{julio@ihes.fr}}, \href{mailto:rtuddenh@uchicago.edu}{\texttt{rtuddenh@uchicago.edu}},

\href{mailto:2051995@tongji.edu.cn}{\texttt{2051995@tongji.edu.cn}}}\par\medskip
\today
\newpage
\tableofcontents

\flushbottom

\section{Introduction}
\label{sec:intro}

An important goal of modern theoretical physics is to understand the space of quantum field theories. This problem is difficult in general, but for \(2+1\)-dimensional topological quantum field theories (TQFTs) it takes a precise algebraic form. The data of a 2+1d TQFT are captured by a modular tensor category, which describes the fusion and braiding of its anyons~\cite{kitaevAnyonsExactlySolved2006}. For a fixed number of anyon types, called the rank, $r,$ there are only finitely many inequivalent modular tensor categories~\cite{bruillardRankfinitenessModularCategories2015}. Finding them is a concrete, if challenging, computational problem. Even the first step, finding their possible modular data, requires searching a space of candidates that grows rapidly with the rank.

In~\cite{Berean-Dutcher:2026fck}, we presented a numerical algorithm with which we identified the bulk of the modular data of modular categories up to rank 22, extending earlier classifications at lower rank~\cite{Rowell:2007dge,Wen:2015qgn,bruillardClassificationModularCategories2015,ngReconstructionModularData2023,ngClassificationModularData2025}. The modular data consist of a pair of matrices $(S,T)$ that define a projective representation of $\sl{\zee{}}$. The matrix $S$ encodes the mutual braiding of anyons, and $T$ is diagonal with entries given by their topological twists. 
\begin{equation}
     S_{ab}=\vcenter{\hbox{\HopfLinkAmplitude{a}{b}}},\qquad T_{ab}=\delta_{ab}\theta_a, ~~\text{for}~~\StandardTwistEquation
\end{equation}
For each candidate representation, the algorithm searches for a basis in which the matrices representing the standard generators of $\sl{\zee{}}$ yield a consistent pair $(S,T)$ after normalization. As the rank grows, the number of candidate representations increases rapidly and a shrinking fraction yield modular data (see \cref{fig:data-by-rank}(b)). 

In this work, we introduce a modified algorithm with three rejection gates that discard unlikely candidates early in the search, thereby extending the reach of our strategy. Two of the gates use logistic regression trained on lower-rank data. All three gates can also discard valid modular data, but exhaustive searches at lower ranks let us measure how many failing representations they reject for a given fraction of modular data retained. At rank 21, for example, cutoffs that retained $80\%$ of the known modular data would have passed fewer than $0.01\%$ of the failing representations through the two regression gates (see \cref{fig:roc}). 

Using this modified algorithm, we have identified 93,287 modular data at ranks 23 through 42. Closing the catalog under known constructions brings the total at these ranks to 150,679. To estimate how much remains to be found, we compare the Chern--Simons theories recovered by the search with those added by closure. Assuming that the gates are unbiased with respect to Chern--Simons theories, we estimate that the catalog contains roughly $80\%$ of the modular data in this range. The data are available at \href{https://AnyonAtlas.org}{\texttt{AnyonAtlas.org}}.

The catalog allows us to test a proposed criterion for time-reversal symmetry in Chern--Simons theories. A TQFT is time-reversal symmetric if it is equivalent to its orientation reversal~\cite{Barkeshli:2016mew,Wang:2016qkb,Tachikawa:2016nmo}; for a Chern--Simons theory with modular category $\mathcal M$, this requires a braided equivalence $\mathcal M\cong\mathcal M^{\text{rev}}$, where $\mathcal M^{\text{rev}}$ is $\mathcal M$ with the braiding reversed. Geiko and Moore~\cite{Geiko:2022qjy} showed that an abelian Chern--Simons theory admits time-reversal symmetry precisely when all its higher Gauss sums are real. They conjectured that the criterion also holds for compact nonabelian gauge groups. We find counterexamples with real higher Gauss sums whose modular data already rule out time-reversal symmetry. The examples suggest a refined criterion for time-reversal symmetry at the level of modular data. A targeted search also yields prime modular data at rank 100 that cannot arise from a Chern--Simons theory with compact gauge group (they arise instead from the Drinfeld center of a near-group fusion category), illustrating the reach of the method beyond the ranks of our systematic survey.

This paper is organized as follows: In \cref{sec:alg} we describe the modified algorithm, and in \cref{sec:data} we describe the catalog through rank 42. We discuss time-reversal symmetry in \cref{sec:tqft} and the rank-100 examples in \cref{sec:100}.

\section{Gated Algorithm}
\label{sec:alg}

\begin{figure}[ht]
    \centering
    \includegraphics{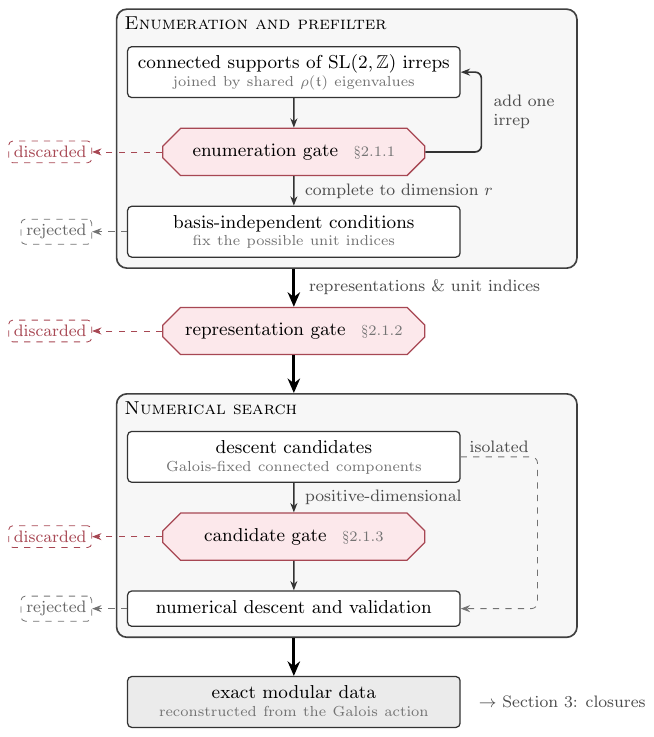}
    \caption{Schematic flowchart of the gated algorithm, indicating the positions of the three gates.}
    \label{fig:flowchart}
\end{figure}

We begin by briefly reviewing the algorithm of~\cite{Berean-Dutcher:2026fck} for identifying modular data, which builds on~\cite{ngReconstructionModularData2023,ngClassificationModularData2025}. Modular data is known to form congruence representations of $\sl{\zee{}}$~\cite{ngCongruenceSubgroupsGeneralized2010}, of which there are only finitely many at a given rank. The algorithm starts from the irreducible congruence representations of $\sl{\zee{}}$, constructed using the classification of~\cite{ngSymmetricRepresentations$textSL_2mathbbZ$2023}, and proceeds in three stages: a prefilter, which forms and prunes candidate representations; a numerical search for a simple-object basis yielding modular data; and, finally, reconstruction of the exact modular data over a cyclotomic field. The modifications described in this section concern the prefilter and the numerical search. A schematic flowchart of the algorithm is presented in \cref{fig:flowchart}, indicating the positions of the three gates.

The algorithm first constructs (reducible) representations whose irreducible summands are connected through shared $\rho(\modt)$ eigenvalues, since only these can yield modular data~\cite{bruillardClassificationModularCategories2015,ngClassificationModularData2025}. The prefilter then tests these representations against basis-independent conditions on modular data, discarding as many failing representations as possible before the costlier numerical search. The strongest of these requirements is that $\rho(\mods)$ admit a basis and a unit index for which no quantum dimension vanishes. This requirement fixes the possible unit indices of a given representation's modular data.

Representations passing all conditions in the prefilter must then be tested for a basis corresponding to modular data. This is a continuous problem which we treat numerically. The search space consists of the changes of basis that keep $\rho(\mods)$ symmetric and $\rho(\modt)$ diagonal,
\begin{equation}\label{eq:changeofbasis}\rho(\mods) = U\tilde\rho(\mods)U^T,~~~~~ U\in\mathrm{Stab}_{O(r)}(\rho(\modt)) \cong \prod_\lambda O(m_\lambda),\end{equation}
where $m_\lambda$ is the multiplicity of the eigenvalue $\lambda$ of $\rho(\modt)$. This space is split into connected lower-dimensional components with fixed Galois action on $(\rho(\mods),\rho(\modt))$, simplifying the problem substantially. We refer to these Galois-fixed connected components, for a given representation, as descent candidates. We then search numerically for a basis in which the matrices yield the correct second Frobenius--Schur indicators and integer fusion coefficients. If such a basis is found, Galois conjugation can be used to consistently reconstruct exact modular data $(S,T)$.

\begin{figure}[ht]
    \centering
    \includegraphics[width=0.95\linewidth]{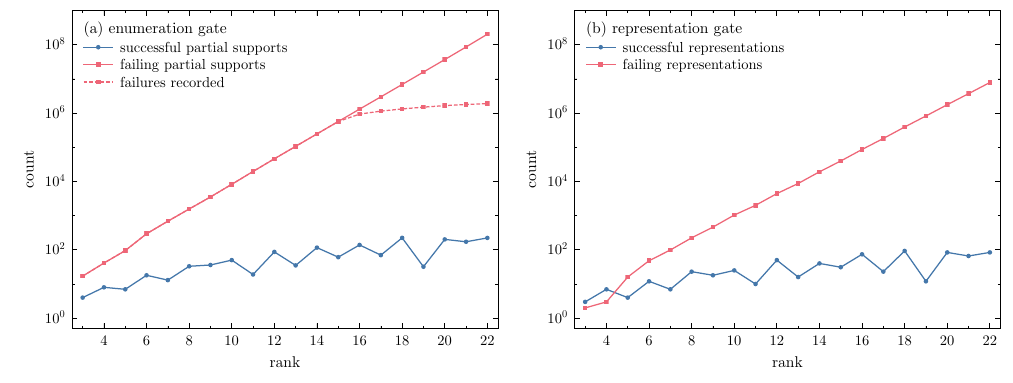}
    \caption{Number of samples considered at each gate, split between those yielding modular data and those that do not. At the enumeration gate~(a), these samples are connected partial supports on the graph of irreducible $\sl{\zee{}}$ representations (see \cref{fig:supports}), while at the representation gate~(b) they are full $\sl{\zee{}}$ representations. At the enumeration gate, only a random fraction of the failing partial supports is recorded beyond rank 15 (dashed line).}
    \label{fig:data-by-rank}
\end{figure}

Although the algorithm described above is effective at identifying new modular data, its runtime grows rapidly with rank, far faster than the number of modular data it identifies. As illustrated in \cref{fig:data-by-rank}, an increasingly small fraction of the representations passing the prefilter actually yield modular data. The rate-limiting step in extending the algorithm to higher rank is enumerating and rejecting this glut of failing representations. We observed, however, that whether a representation or descent candidate is successful correlates strongly with a few of its features. The modification we describe below leverages these correlations to retain and test only the candidates most likely to yield modular data, thereby vastly extending the rank at which new modular data can be efficiently identified. 

We add three gates to the existing algorithm, at which representations unlikely to yield modular data can be rejected. These are implemented as early as practical to avoid wasted computations. First, a logistic regression gate, the enumeration gate, is added during enumeration; it stops constructing an $\sl{\zee{}}$ representation as soon as a score based on its distinct irreducible summands falls below a cutoff. This gate depends closely on how the representations are constructed, which we describe in \cref{sec:enum}. A second logistic regression gate, the representation gate, is implemented after the prefilter, scoring full $\sl{\zee{}}$ representations with candidate unit indices. Finally, the candidate gate applies a fixed rule, based on the Galois action, to discard most of the descent candidates of each representation. The primary obstacle for the two logistic regression gates is that weights and cutoffs determined at known ranks must be applied at a higher rank, where the feature distributions of successful representations are not yet known. This is made more challenging when the data below that rank was also collected with these gates and is therefore incomplete.\footnote{That is, incomplete relative to what the ungated search would have found. The descent itself can also miss modular data~\cite[Sec.~3.3.1]{Berean-Dutcher:2026fck}.} The basic fact that enables our rejection is that certain features continue to correlate strongly with success across all ranks studied.

We emphasize that the aim is not to retain all modular data, which would require giving up most of the rejection. The receiver operating characteristic (ROC) curves in \cref{fig:roc} illustrate this: accepting a small loss of modular data yields a large speed-up, while accepting no loss generally permits little or no rejection at the two logistic regression gates. Some loss is unavoidable at any useful cutoff, since the tails of the feature distributions among representations yielding modular data fluctuate unpredictably. A single loss parameter $\varepsilon$ sets the cutoffs at the two logistic regression gates: they are the optimal pair that would have lost a total fraction $\varepsilon$ of the modular data at lower ranks, extrapolated to the target rank (\cref{sec:eps}). The resulting loss is only an estimate, since the true loss at an unexplored rank cannot be known in advance. 

\subsection{Rejection Gates}

Two of the three gates implement logistic regression~\cite{Cox:1958}, scoring partial supports (on the graph of irreducible $\sl{\zee{}}$ representations) or representations by a set of features found to correlate with yielding modular data. Scores are of the form 
\begin{equation}
    S(x) = c_0 + \sum_j c_j x_j
\end{equation}
for weights $c_j$ and features $x_j$. A partial support or representation is accepted at gate $g$ if $S(x) \geq \tau_g$, where $\tau_g$ is the chosen cutoff for that gate. The fitting of the weights is described in \cref{sec:refit}, and the choice of cutoffs in \cref{sec:eps}. In this section, we describe the features chosen for each gate and how the gates are implemented.

Although each of the two logistic regression gates uses several features, at both gates the weights are dominated by a measure of the number of irreducible summands of the tested support or representation. At every rank, representations with fewer summands are more likely to yield modular data.

\subsubsection{Enumeration Gate}
\label{sec:enum}

\begin{figure}[ht]
    \centering
    \includegraphics{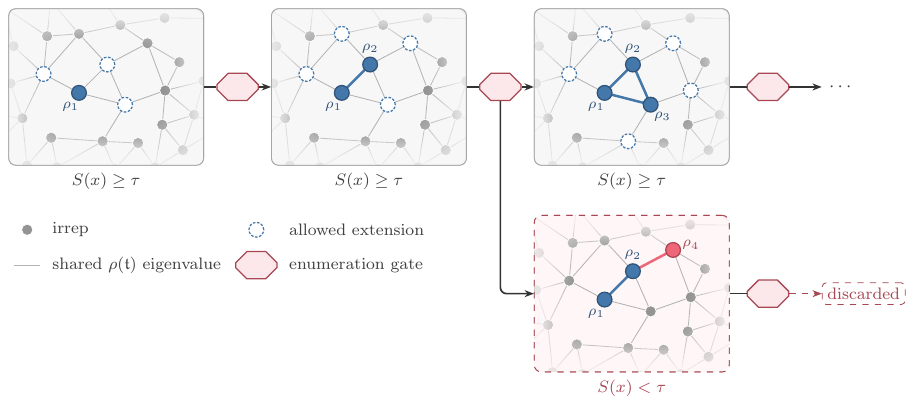}
    \caption{Incremental construction of connected supports on the graph of irreducible $\sl{\zee{}}$ representations, with the enumeration gate applied after each extension.}
    \label{fig:supports}
\end{figure}

The first of the three gates is implemented during the enumeration of congruence representations of $\sl{\zee{}}$. It is positioned there so as to minimize the number of full $\sl{\zee{}}$ representations constructed, since the number of $r$-dimensional congruence representations grows with the rank $r$ as $\exp(r^{2/3+o(1)})$~\cite[App.~B]{Berean-Dutcher:2026fck}. Since only representations whose irreducible summands are connected through shared $\rho(\modt)$ eigenvalues can yield modular data, the enumeration consists of forming all connected supports in the graph whose vertices are the irreducible representations of dimension at most $r$, with an edge between two irreducibles whenever they share a $\rho(\modt)$ eigenvalue. This graphical construction is illustrated in \cref{fig:supports}. We initialize a support at an irreducible, and then extend it one irreducible at a time until multiplicities can be assigned so that the total dimension is $r$. The enumeration gate is applied to these partial supports, scoring them after each incremental extension. 

The nine features used at this gate are as follows, using $\rho = \bigoplus_i \rho_i$ to indicate any representation with the given support $\{\rho_i\}$ and $n=\operatorname{ord}(\rho(\modt))$.
\begin{enumerate}
    \item $|\operatorname{supp} \rho|$: the number of distinct irreducible summands (nodes of the graph) in the partial support.
    \item $\log n_\text{min deg}$: the log level of the lowest-dimensional summand, corresponding to the node on which the support is initialized. Here the level of an irreducible congruence representation $\rho_i$ of $\sl{\zee{}}$ is $n_i = \operatorname{ord}(\rho_i(\modt))$.
    \item $\# \Xi_q/r$: the number of distinct characters in the support, as a fraction of the rank.
    \item $\omega(n)$: the number of distinct prime factors of $n$.
    \item  $\sum_{i<j} |\operatorname{spec}(\rho_i(\modt)) \cap \operatorname{spec}(\rho_j(\modt))|$: the total number of $\rho(\modt)$ eigenvalues shared by pairs of distinct irreducibles in the support, i.e.\ the number of edges among their nodes.
    \item $\nu_2(n)$: the 2-adic valuation of $n$, i.e.\ the exponent of 2 in the prime factorization of $n$ (not to be confused with the Frobenius--Schur indicator $\nu_2$).
    \item $\nu_3(n)$: the 3-adic valuation of $n$.
    \item $d_\text{twist}'$: the sum of the dimensions of distinct twisted summands. Here, twisted refers to the non-character irreducible representations that the \textsf{GAP} package \textsf{SL2Reps}~\cite{ngSymmetricRepresentations$textSL_2mathbbZ$2023} generates in the form $X \otimes \Xi_q$, with $X$ an irreducible representation of any type in the classification of prime-power-level irreducible representations of $\sl{\zee{}}$~\cite{nobsIrreduziblenDarstellungenGruppen1976,nobsIrreduziblenDarstellungenGruppen1976b,ngSymmetricRepresentations$textSL_2mathbbZ$2023} and $\Xi_q$ a nontrivial character. This is a non-unique description, since representations of different types are often related by tensoring with characters.
    \item $d_D'$: the sum of the dimensions of distinct type-$D$ summands in the classification of prime-power-level irreducible representations of $\sl{\zee{}}$~\cite{nobsIrreduziblenDarstellungenGruppen1976,nobsIrreduziblenDarstellungenGruppen1976b,ngSymmetricRepresentations$textSL_2mathbbZ$2023}. These types are determined by Weyl representations arising from distinct quadratic modules.
\end{enumerate}

\begin{figure}[ht]
    \centering
    \includegraphics[width=0.95\linewidth]{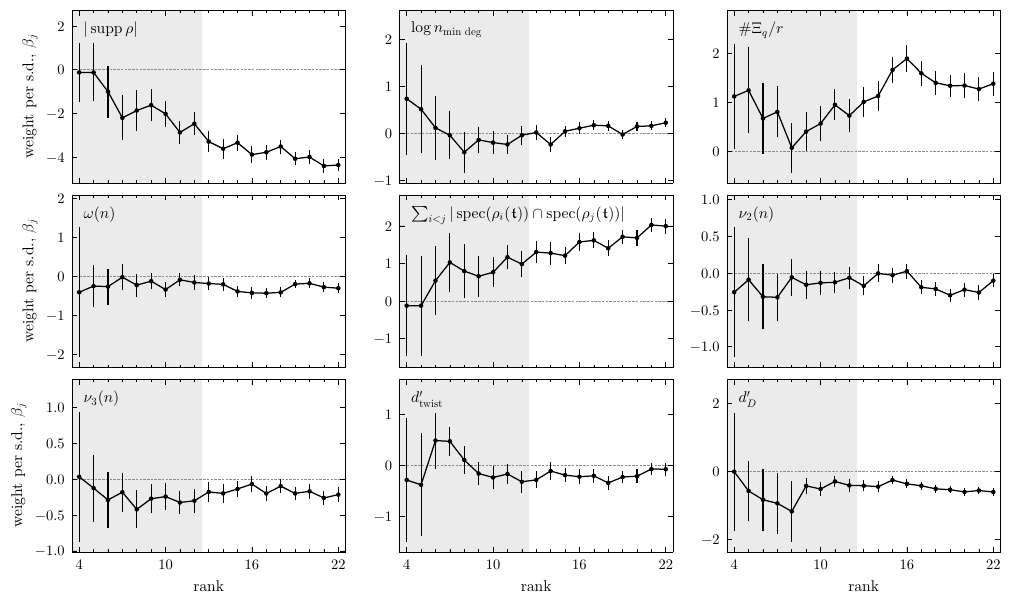}
    \caption{The weights $\beta_j$ (see \cref{sec:refit}) of the enumeration gate features up to rank 22. The largest weight is that of $|\operatorname{supp}\rho|$. Ranks up to 12 are grayed out because the training data there is too sparse to give reliable weights.}
    \label{fig:enum-gate}
\end{figure}

The fitted weights of these features, $\beta_j$ (defined in \cref{sec:refit}), are shown in \cref{fig:enum-gate}. We show $\beta_j$ rather than $c_j$ because the standardized weights do not depend on the scale of the features. The features were selected from a much larger set of candidates because they remain predictive across ranks and are not redundant with one another. The number of distinct irreducible summands, $|\operatorname{supp}\rho|$, carries the largest weight at this gate.


\subsubsection{Representation Gate}

The second gate is simpler than the first, since it tests full congruence representations of $\sl{\zee{}}$, together with the set $U$ of candidate unit indices $u$ of the modular data they might yield. The eight features used at this gate are the following:
\begin{enumerate}
    \item $\log h$: the log of the number of irreducible summands in the representation, counted with multiplicity.
    \item $\frac{1}{r}\sum_i d_i \log n_i$: the  dimension-weighted mean log level, where $n_i = \operatorname{ord}(\rho_i(\modt))$ is the level of the $i$-th summand and $d_i$ is its dimension.
    \item $|U|$: the number of candidate unit indices.
    \item $m_{\mathrm{unit}}=\min_{u\in U}|\{j:\theta_j=\theta_u\}|$: the size of the smallest $\rho(\modt)$-eigenspace containing a candidate unit index $u\in U$.
    \item $f_{\mathrm{eq}}=\frac{\sum_{i<j}\delta_{\theta_i,\theta_j}}{\binom r2}$: the fraction of pairs of twists that are equal. Equivalently, this is the (unreduced) search space dimension of \cref{eq:changeofbasis} as a fraction of the maximum.
    \item $m_{\max}=\max_i|\{j:\rho_j\simeq\rho_i\}|$: the largest multiplicity of an irreducible summand.
    \item $d_R$: the sum of the dimensions of the type-$R$ summands in the classification of prime-power-level irreducible representations of $\sl{\zee{}}$~\cite{nobsIrreduziblenDarstellungenGruppen1976,nobsIrreduziblenDarstellungenGruppen1976b,ngSymmetricRepresentations$textSL_2mathbbZ$2023}. Interestingly, type-$D$ summands correlated more strongly with success when counted without multiplicity, and type-$R$ summands when counted with multiplicity, hence the inclusion of $d_D'$ at the enumeration gate and of $d_R$ at the representation gate.
    \item $|\{i:\tilde\rho(\mods)_{ii} \neq 0\}|$: the number of nonzero diagonal entries of $\tilde\rho(\mods)$. This feature is not invariant under change of basis: it is computed from $\tilde\rho(\mods)$, the matrix $\rho(\mods)$ in the basis produced by \textsf{SL2Reps}~\cite{ngSymmetricRepresentations$textSL_2mathbbZ$2023}, and so suggests a correlation specific to that basis.
\end{enumerate}

\begin{figure}[ht]
    \centering
    \includegraphics[width=0.95\linewidth]{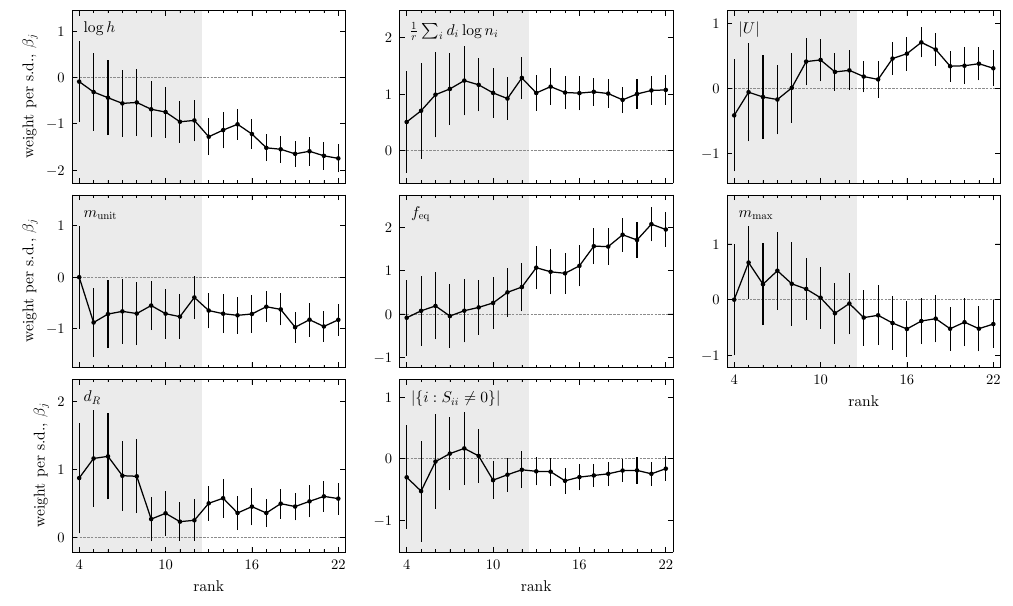}
    \caption{The weights $\beta_j$ (see \cref{sec:refit}) of the representation gate features up to rank 22. Ranks up to 12 are grayed out because the training data there is too sparse to give reliable weights.}
    \label{fig:rep-gate}
\end{figure}

The weights for these features are shown in \cref{fig:rep-gate}. As at the enumeration gate, the number of summands in a representation, here $\log h$, dominates.


\subsubsection{Candidate Gate}

The final gate is the simplest and involves no weights. In the algorithm of~\cite[Sec.~3.2]{Berean-Dutcher:2026fck}, each representation is split into candidates, one for each possible set of signed-permutation matrices $\{D_a\}$~\cite{dongCongruencePropertyConformal2015} by which the Galois group could act on $\rho(\mods)$ and $\rho(\modt)$ in the simple-object basis, if one exists. Fixing the target Galois matrices $\{D_a\}$ restricts the search to the changes of basis that carry the Galois matrices of the initial \textsf{SL2Reps} basis to the $D_a$, which reduces the dimension of the search space. In some cases, the reduced search space is zero-dimensional, so the problem becomes finite and can be solved exactly. Most candidates, however, are still left with positive-dimensional search spaces, and it is to these that the candidate gate is applied.

\begin{figure}[ht]
    \centering
    \includegraphics[width=0.55\linewidth]{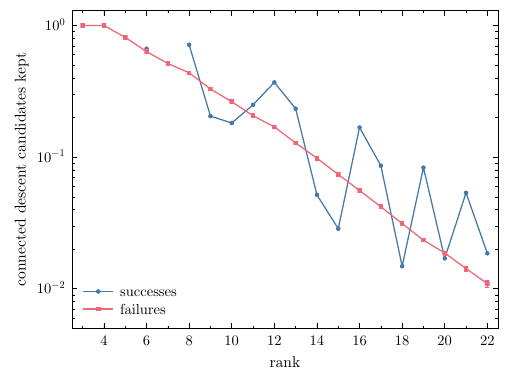}
    \caption{The fraction of connected descent candidates passing the candidate gate, split between representations that yield modular data (successes) and those that do not (failures). This gate does not reject representations, only the connected change-of-basis searches associated with them.}
    \label{fig:cand-gate}
\end{figure}

Each fixed Galois action permutes the simple objects and so partitions them into Galois orbits. By fitting a model to features of the descent candidates, we identified the following pattern: among the descent candidates (having connected, positive-dimensional search spaces) sharing the same representation of $\sl{\zee{}}$, those that yield modular data almost always have the fewest simple-object Galois orbits. Through rank 22, where exhaustive runs exist, only two exceptions are known, each a single representation and choice of Galois action: one at rank 20, yielding \href{https://anyonatlas.org/data/B20-00338}{\texttt{B20-00338}} to \href{https://anyonatlas.org/data/B20-00401}{\texttt{B20-00401}} on \href{https://anyonatlas.org}{\texttt{AnyonAtlas.org}}, and one at rank 22, yielding \href{https://anyonatlas.org/data/B22-01585}{\texttt{B22-01585}} to \href{https://anyonatlas.org/data/B22-01592}{\texttt{B22-01592}}. At the same time, restricting the search to the fixed Galois actions with the fewest simple-object orbits retains a fraction of the descent candidates that falls steeply with rank, to about $1\%$ for failing representations at rank 22, as shown in \cref{fig:cand-gate}. We therefore impose this rule strictly, applying it after Galois reduction to the positive-dimensional descent candidates.

\subsection{Logistic Model} 
\label{sec:refit}

The logistic models at the enumeration and representation gates are fitted by $L_2$-regularized logistic regression, with rank-dependent intercepts and class-balanced sample weights. The training set at a given rank $r$ consists of the samples\footnote{We use the term sample for a partial support at the enumeration gate and for a full $\sl{\zee{}}$ representation at the representation gate.} recorded in previous (possibly gated) runs at ranks below $r$. Refitting at each rank allows the weights to vary with rank, but training on gated runs, which contain false negatives, can cause the weights to drift.\footnote{This could be mitigated substantially by fitting the weights after performing the closures described in \cref{sec:data}. Since the closure is involved and the drift contributes less to the loss of modular data than the choice of cutoffs, we did not close the data incrementally in this way.} This is unavoidable with the current design, and none of the extrapolations of trends from ranks below $r$ that we tested performed better than this simple refitting. For the representation gate, we can easily store all failing and successful representations; those already rejected by the prefilter are excluded from training. For the enumeration gate, failing partial supports are so numerous that we record only a random sample of them (see \cref{fig:data-by-rank}). With $(i)$ indexing the samples in the training set, we first convert to standardized features
\begin{equation}
    z_j^{(i)} = \frac{x_j^{(i)} - \mu_j}{\sigma_j}
\end{equation}
where $\mu_j$ and $\sigma_j$ are the mean and standard deviation of $x_j$ across the full training set. Note that three of the features depend explicitly on the rank of the sample $(i)$. The linear predictor
\begin{equation}
    \eta^{(i)} = \alpha_{r_i} + \sum_j \beta_j z_j^{(i)}
\end{equation}
involves, unlike standard logistic regression, a separate intercept $\alpha_{r_i}$ for each training rank, where $r_i$ is the rank of sample $(i)$. These rank-dependent intercepts account for the rapid decrease with rank in the probability that a sample with given feature values yields modular data (cf.\ \cref{fig:data-by-rank}). For the purposes of fitting, a sample is a success, labeled $y^{(i)}=1$, if it yields modular data, and a failure, labeled $y^{(i)}=0$, otherwise. At the enumeration gate, a partial support is successful if at least one representation constructed from it yields modular data. We comment on other measures of success at the end of this section.

The feature weights $\beta_j$ and intercepts $\alpha_{r_i}$ are fitted by minimizing the weighted, $L_2$-penalized negative log-likelihood
\begin{equation}\label{eq:likelihood}
    L(\alpha,\beta) = \sum_i w^{(i)} \left[ \log\left(1+e^{\eta^{(i)}}\right) - y^{(i)} \eta^{(i)}\right] + \frac{\lambda}{2} \sum_{j} \beta_j^2
\end{equation}
The sample weights $w^{(i)}$ serve several purposes. First, we balance them as 
\begin{equation}
    \sum_{i:\,y^{(i)} = 1} w^{(i)} = \sum_{i:\,y^{(i)} = 0} w^{(i)} = P \coloneqq |\{i: y^{(i)} = 1\}|
\end{equation}
so that the vast number of negative samples does not drown out the successes. Second, at the enumeration gate, the weights rescale the contribution of the randomly sampled failing partial supports. Finally, they can be used to weight some successes more heavily than others. For example, suppose one wished to maximize the total number of distinct modular data identified. Then, with $m^{(i)}$ the number of modular data yielded by a successful sample $(i)$ at the representation gate and $N = |\{i: y^{(i)} = 0\}|$ the number of failing samples, one could choose the weights
\begin{equation}
    w^{(i)} = \begin{cases}
        m^{(i)}\,\frac{P}{\sum_{k:\,y^{(k)}=1} m^{(k)}}, & y^{(i)} = 1\\
        \frac{P}{N}, & y^{(i)} = 0
    \end{cases}
\end{equation}
The last term in \cref{eq:likelihood} is an $L_2$ (ridge) penalty on large weights, with strength $\lambda \geq 0$, needed to account for the sparsity of the training data and for correlations between features.\footnote{We have also found that the model is improved by adding a penalty $\sum_{r'<r}\alpha_{r'}^2$ on the intercepts, with a very small coefficient.} We use $\lambda = 1/3$ at the enumeration gate and $\lambda = 1$ at the representation gate.

\begin{figure}[ht]
    \centering
    \includegraphics[width=0.95\linewidth]{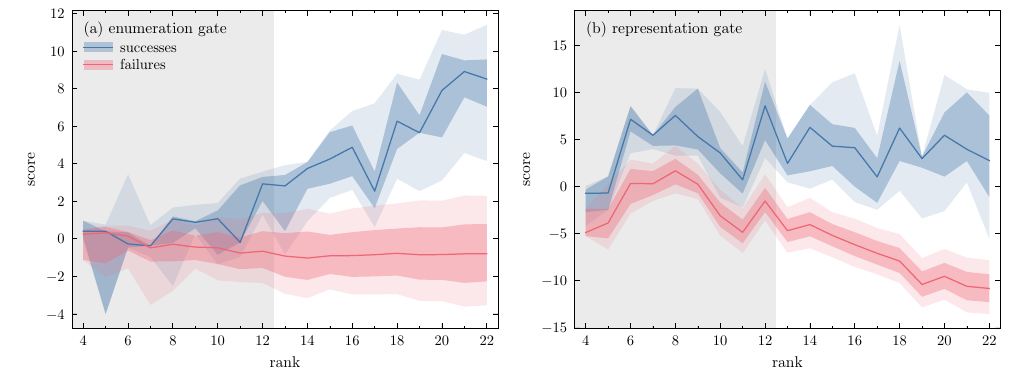}
    \caption{Scores at (a) the enumeration gate and (b) the representation gate, split between samples that do and do not yield modular data. Ranks up to 12 are grayed out, as in \cref{fig:enum-gate,fig:rep-gate}.}
    \label{fig:scores}
\end{figure}

Finally, given the fitted weights, a gated run at rank $r$ assigns each sample the score
\begin{equation}
    S(x) = \sum_j\beta_j\,\frac{x_j - \mu_j}{\sigma_j}
\end{equation}
The intercepts $\{\alpha_{r'}\}_{r'<r}$ do not enter the score: they do little to predict the intercept $\alpha_r$ at rank $r$, and in any case a constant shared by all samples at rank $r$ shifts every score equally and can be absorbed into the cutoff. In terms of the unstandardized features, the score takes the form
$S(x) = c_0 + \sum_j c_j x_j$
for
\begin{equation}
    c_j = \frac{\beta_j}{\sigma_j},\qquad c_0 = -\sum_j \frac{\beta_j \mu_j}{\sigma_j}
\end{equation}
Scores at the two gates up to rank 22 are shown in \cref{fig:scores}, split between failing and successful samples. Each gate $g$ rejects a sample if its score falls below the cutoff $\tau_g$. These cutoffs, which determine the speed-up over the exhaustive search, are chosen separately from the fit, as described in \cref{sec:eps}.

We also tested the logistic regression model with other definitions of success. Of most interest is modular data that cannot be obtained by the constructions of \cref{sec:data}, for example prime modular data that does not correspond to a connected Chern--Simons or Dijkgraaf--Witten theory under any change in spherical structure. Targeting such exotic modular data did not produce meaningfully stronger correlations with any of the features tested, while the sparser training data made the cutoffs described in the following section less predictable. Taking all modular data as successes therefore appears to give the best results.

\subsection{Cutoffs}
\label{sec:eps}

\begin{figure}[ht]
    \centering
    \includegraphics[width=\linewidth]{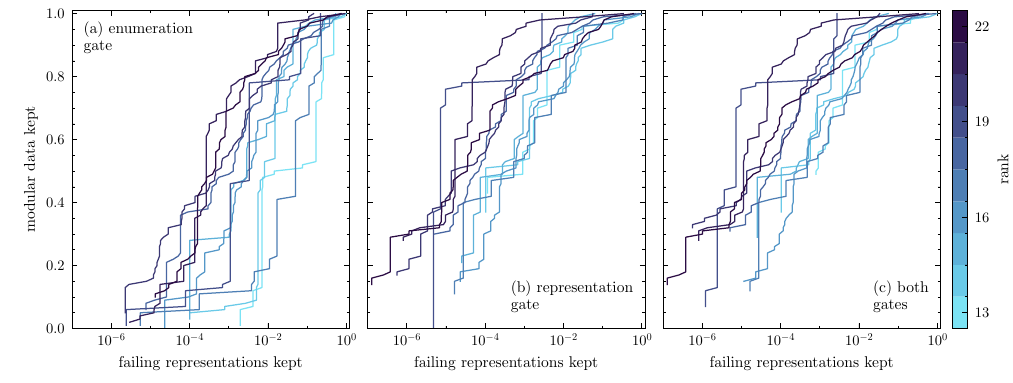}
    \caption{ROC curves for (a) the enumeration gate and (b) the representation gate at ranks 13 to 22, and (c) joint ROC curves for the optimal split of cutoffs between the two gates. Each curve shows, at a given rank, the fraction of failing representations kept ($x$-axis) against the fraction of known modular data kept ($y$-axis).}
    \label{fig:roc}
\end{figure}

\begin{figure}[ht]
    \centering
    \includegraphics[width=0.6\linewidth]{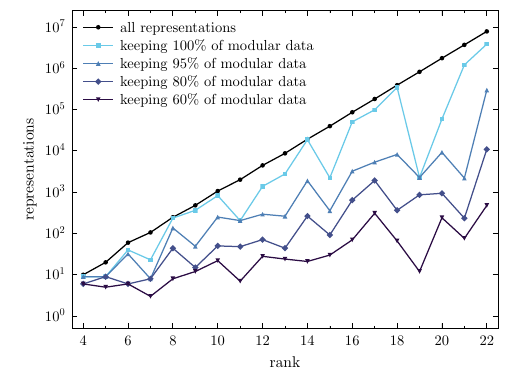}
    \caption{Number of $\sl{\zee{}}$ representations that must be searched to identify a given fraction of the modular data. Although this number fluctuates considerably from rank to rank, even a small loss gives a large reduction in runtime.}
    \label{fig:runtime}
\end{figure}

The runtime improvements from these logistic regression gates come from selecting score cutoffs that lose some fraction of the modular data but reject a much larger portion of failing representations. This is illustrated by the ROC curves in \cref{fig:roc}, which show, for ranks 13 to 22 (where exhaustive runs exist and the fitted weights are reliable), the fraction of failing representations that must be searched to recover a given fraction of the modular data. For example, to recover $80\%$ of the known modular data at rank 21, we could have set cutoffs keeping less than $0.01\%$ of the failing representations across the two gates. Importantly, the ROC curves show that the rejection tends to become stronger with rank, partly offsetting the rapid growth of the problem: as rank grows, a smaller fraction of the failing representations needs to be searched to recover the same fraction of the modular data. Combining the ROC curves with the representation counts of \cref{fig:data-by-rank} gives the number of representations that must be searched at each rank for a given loss fraction, shown in \cref{fig:runtime}.

Estimating the fraction of modular data recovered at higher rank, however, requires some knowledge of the score distributions of as yet unidentified data. Our strategy for this is somewhat approximate, so we only outline it schematically; it is nonetheless sufficient to identify modular data reliably. Given a target loss fraction $\varepsilon$ at rank $r$, we evaluate the scores at reference ranks where exhaustive runs have been completed (ranks up to 22 in our case) and determine the runtime-optimal cutoffs at the two gates that would have lost a fraction $\varepsilon$ of the data. At the enumeration gate, a representation is constructed only if every partial support leading to it passes, so each representation is assigned the minimum score over these partial supports. We then extrapolate these cutoffs to rank $r$ using fits that account for the spread of scores in the gated runs at ranks between 22 and $r$. In principle, the loss at rank $r$ could instead be estimated from the constructible modular data at that rank, as in the completeness estimate of \cref{sec:data}. We have not attempted this, as it could bias the cutoffs toward constructible, and hence less interesting, data.

\section{Modular Data through Rank 42}
\label{sec:data}

\begin{figure}[ht]
    \centering
    \includegraphics[width=0.65\linewidth]{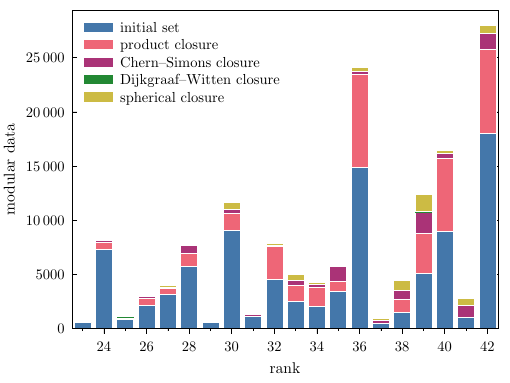}
    \caption{Counts of modular data identified between ranks 23 and 42 by source: the gated search (initial set) and the product, Chern--Simons, Dijkgraaf--Witten and spherical closures.}
    \label{fig:sources}
\end{figure}

Thus far, the gated search of \cref{sec:alg} has identified 93,287 modular data between ranks 23 and 42. The search was run for around three weeks on a dual-socket AMD EPYC 9654 machine provided by \href{https://projectnumina.ai}{Project Numina}. By design, the gates lose some modular data, so the search output is incomplete. We bring this set closer to completeness by closing it under several known constructions:
\begin{enumerate}
    \item Products: Deligne products $\mathcal M_1 \boxtimes \mathcal M_2$ are formed from the lower-rank modular data identified in~\cite{Berean-Dutcher:2026fck}. This adds 39,854 non-prime data to the catalog between ranks 23 and 42.
    \item Connected Chern--Simons Theories: Using a modified version of the Chern--Simons calculator of~\cite{Berean-Dutcher:2026fck}, which implements the simple-current extensions of~\cite{Fuchs:1996dd}, we compute all invertible condensations from categories of the form
    \begin{equation}
        \mathcal P \boxtimes \C(\mathfrak g_1,k_1) \boxtimes \dots \boxtimes \C(\mathfrak g_n,k_n)
    \end{equation}
    of rank up to 3500. Here, $\mathcal P$ is any pointed modular category, described equivalently as a metric group category $\mathcal C(A,q)$, with $A$ a finite abelian group and $q$ a non-degenerate quadratic form, or as the line category of a $\Uone^r_K$ Chern--Simons theory. Similarly, the affine Lie algebra category $\C(\mathfrak g,k)$ is the line category of the level-$k$ Chern--Simons theory whose gauge group is the simple, simply connected group with Lie algebra $\mathfrak g$. Since quotients of the gauge group by central subgroups are realized by condensing invertible bosons~\cite{mooreTamingConformalZoo1989}, invertible condensations of these categories describe all Chern--Simons theories with compact connected gauge group. We use these condensates both to identify Chern--Simons theories among the data already found and, together with their Galois conjugates, to add 10,480 further modular data.
    \item Dijkgraaf--Witten Theories: We add the Dijkgraaf--Witten theories~\cite{dijkgraafTopologicalGaugeTheories1990} not already obtained by the above closures, contributing only 107 further modular data after closing under Galois conjugation.
    \item Spherical Structure: Finally, we close the data under changes of spherical structure~\cite{bruillardRankfinitenessModularCategories2015,bruillardClassificationModularCategories2015}, which multiply $d_a$ and $\theta_a$ by $\chi(a)$ for a $\pm1$-valued character $\chi$ of the universal grading group, adding 6,951 further modular data.
\end{enumerate}
Together with the 93,287 data found by the search, this gives 150,679 modular data between ranks 23 and 42, bringing the number of bosonic modular data available at \href{https://AnyonAtlas.org}{\texttt{AnyonAtlas.org}} to 169,011. \Cref{fig:sources} shows, at each rank, how many data come from the search and from each closure. Of the data found by the search, 7,699 are identified as connected Chern--Simons theories, while a further 7,486 were added by the product and Chern--Simons closures. If the gates are biased neither toward nor against Chern--Simons theories (as is the case through rank 22), this suggests that the search alone found $\approx 50\%$ of the modular data at these ranks.\footnote{Note that in both cases here we only identify Chern-Simons theories by condensation up to rank 3,500, and others may exist both in the initial data and beyond the current closure.} Since the closures enlarge the search output by over $60\%$, we estimate, somewhat roughly, that around $80\%$ of the modular data between ranks 23 and 42 has been identified.

\begin{figure}[ht]
    \centering
    \includegraphics[width=\linewidth]{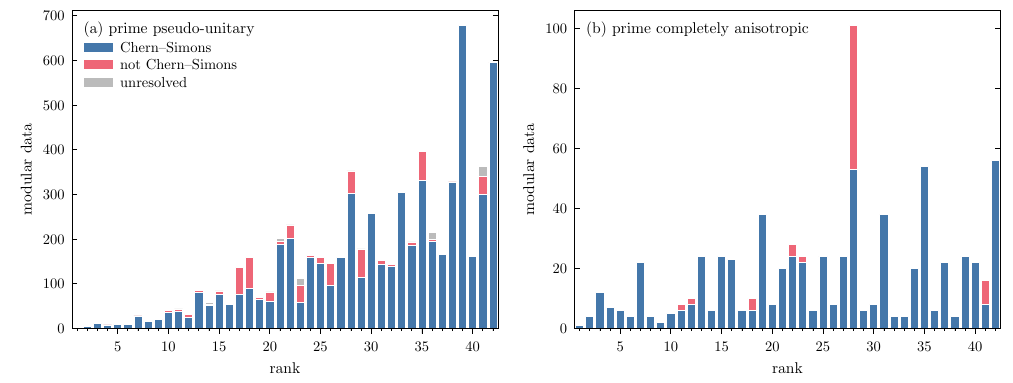}
    \caption{Counts of prime pseudo-unitary modular data through rank 42, (a) all and (b) completely anisotropic, split into Chern--Simons theories, data that are provably not Chern--Simons, and unresolved data, for which no Chern--Simons realization has been found but none has been ruled out. Panel (b) shows only data that can be proven anisotropic from the current catalog. A further 918 prime data may be anisotropic: 874 are Chern--Simons, 4 are proven not to be, and 40 are unresolved but likely arrises $H$-gaugings of connected Chern--Simons theories.}
    \label{fig:cs}
\end{figure}

Lastly, we count the prime pseudo-unitary data that are, and those that provably are not, realized by a Chern--Simons theory with compact gauge group. As in~\cite[Sec.~4.2.1]{Berean-Dutcher:2026fck}, we count as Chern--Simons the gaugings of 0-form $H$-symmetries of connected Chern--Simons theories~\cite{barkeshliSymmetryFractionalizationDefects2019}, which account for disconnected gauge groups $G$ with $\pi_0(G) = H$ (in particular all Dijkgraaf--Witten theories). The resulting counts are shown in \cref{fig:cs}(a). \Cref{fig:cs}(b) restricts to the completely anisotropic data. By the arguments of~\cite{Berean-Dutcher:2026fck}, those without a Chern--Simons realization, if categorifiable, likely represent Witt classes~\cite{davydovWittGroupNondegenerate2011} with no Chern--Simons representative; this would be guaranteed by an affirmative answer to the question posed in~\cite[Sec.~5]{Berean-Dutcher:2026fck}, which remains consistent with the data through rank 42.

\section{Applications}
\label{sec:app}

\subsection{Time Reversal in 2+1d TQFT}
\label{sec:tqft}

We hope that the data we have collected will enable a better understanding of modular categories and their physical counterparts. As an example of this, we test and refine a conjecture of~\cite{Geiko:2022qjy} on necessary and sufficient conditions for time-reversal symmetry in Chern--Simons theories~\cite{Barkeshli:2016mew,Wang:2016qkb,Tachikawa:2016nmo}. We identify a Chern--Simons theory with its modular category $\mathcal M$; time-reversal symmetry then requires a braided equivalence
\begin{equation}
    \mathcal M \cong \mathcal M^{\text{rev}}
\end{equation}
At the level of modular data, such an equivalence induces a permutation $\pi$ of the simple objects that fixes the unit and satisfies
\begin{equation}
    S_{\pi(a),\pi(b)} = S_{ab}^*, \qquad \theta_{\pi(a)} = \theta_a^*
\end{equation}
The existence of such a permutation is necessary but not in general sufficient, since $\pi$ need not lift to a braided equivalence. For pointed modular categories (abelian Chern--Simons theories), Geiko and Moore~\cite{Geiko:2022qjy} prove that the reality of the higher Gauss sums~\cite{ngHigherGaussSums2019}
\begin{equation}
\tau_n(\mathcal M)
=\sum_{a\in\irr(\mathcal M)}d_a^2\theta_a^n,
\qquad 0<n <\operatorname{ord}(T)
\end{equation}
is a necessary and sufficient condition for the theory to be time-reversal symmetric. 
That the reality of the $\tau_n$ is necessary for time-reversal symmetry follows directly from their definition, but it can also be understood physically. The $\tau_n$ can be recovered from the action of the mapping class group $\sl{\zee{}}$ on the torus Hilbert space as
\begin{equation}
    \bra{1} ST^nS\ket{1} = \frac{\tau_n}{D^2}
\end{equation}
By surgery on the $n$-framed unknot, this matrix element equals the partition function $\mathcal Z_{\mathcal M}(M)$ of the lens space $M = L(n,1)$ up to framing. Then, in a unitary theory, time-reversal symmetry requires that $\mathcal Z_{\mathcal M}(\overline{M}) = \mathcal Z_{\mathcal M}(M)^* = \mathcal Z_{\mathcal M}(M)$ and so the $\tau_n$ must be real. It is not obvious that these lens-space conditions are sufficient, but in the pointed case they are.

It is further conjectured in~\cite{Geiko:2022qjy} that the reality of the higher Gauss sums is sufficient for time-reversal symmetry, at least in Chern--Simons theories with compact gauge group. Our data show that this conjecture is false. Specifically, 652 of our modular data have real $\tau_n$ for all $n$ but admit no permutation $\pi$ as above. Of these, 38 are pseudo-unitary, all of which are realized by a Chern--Simons theories with compact gauge group, providing counterexamples to the conjecture. The lowest-rank example, at rank 14, is \href{https://anyonatlas.org/data/B14-00086}{\texttt{B14-00086}}, realized by $(\Uone^5_K\times \SU(3)_5)/\zee{3}$ with $K = -\operatorname{tridiag}(1,2,1)$. There are also examples among Chern--Simons theories with simply connected gauge group, \href{https://anyonatlas.org/data/B36-22062}{\texttt{B36-22062}}: $\Spin(5)_1^2 \times \Spin(14)_1$, and among Dijkgraaf--Witten theories, \href{https://anyonatlas.org/data/B22-00134}{\texttt{B22-00134}}: $\operatorname{DW}(Q_8,\omega) \cong (\SU(2)_2^3\times \Spin(7)_1)/\zee{2}$. We emphasize that, although we have only the modular data of these examples, their explicit realizations guarantee that categorifications exist.

The data suggest a refined conjecture involving the $\nu_2$-weighted Gauss sums
\begin{equation}
    \hat \tau_n(\mathcal M) = \sum_{a\in\irr(\mathcal M)} \nu_2(a)\, d_a^2 \theta_a^n,
\end{equation}
where $\nu_2(a)$ is the second Frobenius--Schur indicator of $a$~\cite{ngHigherFrobeniusSchurIndicators2007}. 
\begin{conjecture}\label{conj:time-reversal}
A pseudo-unitary modular category $\mathcal M$ has time-reversal symmetry at the level of its modular data if and only if both $\tau_n(\mathcal M)$ and $\hat\tau_n(\mathcal M)$ are real for every integer $1\leq n<\operatorname{ord}(T)$.
\end{conjecture}
This condition is clearly necessary, since $\nu_2$ is preserved under time reversal. Given the role of $\nu_2$ as the orientation-reversal anomaly of a self-dual line, its appearance here is natural, but we are not aware of any deeper significance of the particular form of $\hat\tau_n$. In pseudo-unitary pointed categories, reality of the $\tau_n$ implies reality of the $\hat\tau_n$\footnote{In a pointed category, $\nu_2(a) = d_a\theta_a^{-2}$ if $a$ is self-dual and $\nu_2(a) = 0$ otherwise.} (although the converse is not true; for example, $\SU(3)_1$ has $\hat\tau_n = 1$ for all $n$ but $\tau_1 = i\sqrt3$), so these are not independent conditions. For non-pseudo-unitary data, reality of the $\tau_n$ and $\hat\tau_n$ is not sufficient, but in the known data the requirement that the multiset of triples
\begin{equation}
    \left\{(d_a,\theta_a,\nu_2(a)):a\in\irr(\mathcal M)\right\}
\end{equation}
be invariant under complex conjugation is sufficient.

Interestingly, in pseudo-unitary cases, real higher Gauss sums appear to be sufficient for a form of non-invertible time-reversal symmetry, for which only the reality of the $\tau_n$ with $\gcd(n,\operatorname{ord}(T))=1$ is necessary~\cite{ngHigherGaussSums2019}. Such symmetries have been studied in, for example,~\cite{Cordova:2023bja}. For our purposes, we define non-invertible time-reversal symmetry to mean that there exists a topological interface between $\mathcal M$ and $\mathcal M^\text{rev}$. In terms of Witt equivalence classes~\cite{davydovWittGroupNondegenerate2011}, this means that
\begin{equation}
    [\mathcal M] = [\mathcal M^\text{rev}]
\end{equation}
or, equivalently, $[\mathcal M]^2 = 1$. All pseudo-unitary cases in the known data whose $\tau_n$ are real but whose $\hat\tau_n$ are not real possess this non-invertible form of time-reversal symmetry.

\subsection{Modular Data at Rank 100}
\label{sec:100}

To demonstrate the reach of the gated search, which identifies high-rank modular data from features correlated with success at lower ranks, we exhibit prime modular data at rank 100 that does not correspond to a Chern--Simons theory. Although exhaustive scans become increasingly costly with rank, extending the search to very high rank presents no insurmountable obstacle if one seeks only new examples rather than a complete classification. A more targeted gated search at rank 100 leveragin the correlations described above identified numerous modular data. While most of these corresponded to Chern--Simons theories with known constructions, several did not, and we describe a pair of them here.

The modular data $\mathcal M_{a,b}$ we find depends on a pair of integers $(a,b)$, fixed below, and has simple objects ordered as
\begin{equation}
    \irr(\mathcal M_{a,b}) = \{1,B,P_1,\dots,P_{48}, Q_1,\dots, Q_{50}\}
\end{equation}
For
\begin{equation}
    x = \frac{1}{2}\left(97 + \sqrt{9797}\right)
\end{equation}
its total quantum dimension is $D = \sqrt{97}(x + 2)$. Its $T$-matrix is
\begin{equation}
    T = \operatorname{diag}\left(1,1,\left\{\zeta_{97}^{ai^2}\right\}_{i=1}^{48}, \left\{\zeta_{101}^{bj^2}\right\}_{j=1}^{50}\right)
\end{equation}
where $\zeta_n = e^{2\pi i /n}$. In terms of the matrices
\begin{align}
    A^{(a)}_{ij}
&=2\cos\left(\frac{4\pi aij}{97}\right),
~~~~~1\le i,j\le48\\
B^{(b)}_{ij}
&=2\cos\left(\frac{4\pi bij}{101}\right),
~~~~~1\le i,j\le50
\end{align}
its $S$-matrix takes the block form
\begin{equation}
    S=\frac1D
\begin{pmatrix}
1 & x+1 & (x+2)\mathbf1_{48}^{\top} & x\mathbf1_{50}^{\top}\\
x+1 & 1 & (x+2)\mathbf1_{48}^{\top} & -x\mathbf1_{50}^{\top}\\
(x+2)\mathbf1_{48} & (x+2)\mathbf1_{48} & (x+2)A^{(a)} & 0\\
x\mathbf1_{50} & -x\mathbf1_{50} & 0 & -xB^{(b)}
\end{pmatrix}
\end{equation}
Here $\mathbf1_k$ is the all-ones column vector of length $k$. The above $(S,T)$ yield pseudo-unitary modular data satisfying all known conditions (see~\cite[Secs.~3.1 and~3.3.4]{Berean-Dutcher:2026fck}) for $(a,b) = (1,2)$ and $(5,1)$; the two data are Galois conjugates. The former has chiral central charge $c_- \equiv 0 \pmod 8$, while the latter has $c_- \equiv 4 \pmod 8$.

By the dimension-field argument of~\cite[Sec.~4.2.1]{Berean-Dutcher:2026fck}, this modular data is not realized by any Chern--Simons theory with compact gauge group: its dimension field $\mathbb Q(D^2) = \mathbb Q(\sqrt{9797})$ is not a compositum of the affine Lie dimension fields identified in~\cite{gannonAlgebraicNumberFields2019}. Although not described by a Chern--Simons theory, this particular modular data is known, originating from the center of an Evans--Gannon near-group fusion category of type $G + |G|$~\cite{Evans:2012ta,Grossman_2020}. These are fusion categories $\mathcal N$ for a finite abelian group $G$ of odd order $n$, with
\[\irr(\mathcal N) = G\cup \{\rho\}\]
and fusion rules
\[
\begin{aligned}
g\otimes h&=g+h,\\
g\otimes\rho&=\rho\otimes g=\rho,\\
\rho\otimes\rho&=\bigoplus_{g\in G}g\oplus n\rho.
\end{aligned}
\]
Constructions of this kind were recently shown to be categorifiable~\cite{radchenko2026realquadraticfieldsfinite,schopieray2026stretchedneargroups}. The Drinfeld center of $\mathcal N$ contains a nondegenerate pointed subcategory associated with a quadratic form on $G$, and consequently it admits a decomposition
\[
\mathcal Z(\mathcal N)
\simeq\mathcal C(G,q_0)\boxtimes\mathcal M
\]
Here, $\mathcal M$ has rank $n+3$. Factorizations of this form for near-group categories with $G = \zee{97}$ yield the identified modular data.

\begin{equation}
    \C\left(\zee{97},\frac{a y^2}{97}\right) \cong \begin{cases}
        \SU(97)_1, & a=1\\ \operatorname{U}(1)_K^4,~~~K=\left(\begin{smallmatrix}
            2 & -1 & 0 & 0\\-1 & 4 & -1 & 0\\0 & -1 & 4 & -1\\0 & 0 & -1 & 4
        \end{smallmatrix}\right), & a=5
    \end{cases}
\end{equation}
Since $\mathcal M\boxtimes\mathcal C(G,q_0)\simeq\mathcal Z(\mathcal N)$ is Witt-trivial, $\mathcal M$ is Witt-equivalent to a pointed anisotropic category on $\zee{97}$:
\begin{equation}
    \C\left(\zee{97},\frac{a y^2}{97}\right) \cong \begin{cases}
        \SU(97)_1, & a=1\\ \operatorname{U}(1)_K^4,~~~K=\left(\begin{smallmatrix}
            2 & -1 & 0 & 0\\-1 & 4 & -1 & 0\\0 & -1 & 4 & -1\\0 & 0 & -1 & 4
        \end{smallmatrix}\right), & a=5
    \end{cases}
\end{equation}
$\mathcal M$ therefore contains a connected étale algebra $A = 1 + B$ whose local modules give that category~\cite{davydovWittGroupNondegenerate2011}. Both instances possess time-reversal symmetry at the level of their modular data, given by the permutation
\begin{align}
\pi(1)&=1,&
\pi(B)&=B,\\
\pi(P_i)&=P_{[22i]_{97}},&
\pi(Q_j)&=Q_{[10j]_{101}}
\end{align}
where $[k]_p$ denotes the unique element of $\{1,\dots,(p-1)/2\}$ congruent to $\pm k$ modulo $p$.

\FloatBarrier

\vspace{5pt}
\noindent\textbf{Acknowledgements.}
We are grateful to \href{https://projectnumina.ai}{Project Numina} for computing support and collaboration on related topics. JBD is supported by a Four-Year Fellowship (4YF) from the University of British Columbia. CC and RT are supported by the Simons Collaboration on Global Categorical Symmetries and the US Department of Energy Grant 5-29073. The gates and logistic model were designed by the authors, while their implementation benefited from the use of AI models. 


\printbibliography

\end{document}

%% file: figures/tikz-diagrams/ribbon-twist-diagrams.tex
\input{figures/tikz-diagrams/diagram-style.tex}
\input{figures/tikz-diagrams/ribbon-geometry.tex}

\newcommand{\StandardTwistSelfLoop}[1]{%
  \begin{tikzpicture}[baseline=-.52ex,x=.48cm,y=.48cm]
    \path[use as bounding box] (-2.15,-2.4) rectangle (.4,2.4);
    \DrawRibbonCenter{appendix strand}
    \DrawRibbonCenterOverpass{appendix crossing}
    \node[appendix label,above=2pt] at (0,2) {$#1$};
    \node[appendix label,below=2pt] at (0,-2) {$#1$};
  \end{tikzpicture}%
}
\newcommand{\StandardStraightLine}[1]{%
  \begin{tikzpicture}[baseline=-.52ex,x=.48cm,y=.48cm]
    \path[use as bounding box] (-.43,-2.4) rectangle (.43,2.4);
    \draw[appendix strand] (0,-2) -- (0,2);
    \node[appendix label,above=2pt] at (0,2) {$#1$};
    \node[appendix label,below=2pt] at (0,-2) {$#1$};
  \end{tikzpicture}%
}
\newcommand{\StandardTwistEquation}{%
  \vcenter{\hbox{\StandardTwistSelfLoop{a}}}
  =\theta_a\,\vcenter{\hbox{\StandardStraightLine{a}}}%
}

%% file: figures/tikz-diagrams/diagram-style.tex
\tikzset{
  appendix strand/.style={
    draw=black,
    line width=0.75pt,
    line cap=round,
    line join=round
  },
  appendix crossing/.style={
    appendix strand,
    preaction={draw=white, line width=3.2pt}
  },
  appendix vertex/.style={circle, fill=black, inner sep=1.25pt},
  appendix label/.style={font=\scriptsize, inner sep=1pt},
  appendix channel/.style={font=\scriptsize, inner sep=0pt},
  appendix basis/.style={font=\scriptsize, inner sep=0pt}
}

%% file: figures/tikz-diagrams/ribbon-geometry.tex
\newcommand{\DrawRibbonCenter}[1]{%
    \draw[#1] (0.000000,-2.000000)
      .. controls (0.000000,-1.750000) and (-0.093000,-1.512210) .. (-0.210000,-1.255800)
      .. controls (-0.231667,-1.208317) and (-0.253125,-1.161971) .. (-0.274375,-1.116741)
      .. controls (-0.295625,-1.071510) and (-0.316667,-1.027396) .. (-0.337500,-0.984375)
      .. controls (-0.358333,-0.941354) and (-0.378958,-0.899427) .. (-0.399375,-0.858572)
      .. controls (-0.419792,-0.817717) and (-0.440000,-0.777933) .. (-0.460000,-0.739200)
      .. controls (-0.480000,-0.700467) and (-0.499792,-0.662783) .. (-0.519375,-0.626128)
      .. controls (-0.538958,-0.589473) and (-0.558333,-0.553846) .. (-0.577500,-0.519225)
      .. controls (-0.596667,-0.484604) and (-0.615625,-0.450990) .. (-0.634375,-0.418359)
      .. controls (-0.653125,-0.385729) and (-0.671667,-0.354083) .. (-0.690000,-0.323400)
      .. controls (-0.708333,-0.292717) and (-0.726458,-0.262996) .. (-0.744375,-0.234216)
      .. controls (-0.762292,-0.205435) and (-0.780000,-0.177596) .. (-0.797500,-0.150675)
      .. controls (-0.815000,-0.123754) and (-0.832292,-0.097752) .. (-0.849375,-0.072647)
      .. controls (-0.866458,-0.047542) and (-0.883333,-0.023333) .. (-0.900000,0.000000)
      .. controls (-0.916667,0.023333) and (-0.933125,0.045792) .. (-0.949375,0.067397)
      .. controls (-0.965625,0.089002) and (-0.981667,0.109754) .. (-0.997500,0.129675)
      .. controls (-1.013333,0.149596) and (-1.028958,0.168685) .. (-1.044375,0.186966)
      .. controls (-1.059792,0.205246) and (-1.075000,0.222717) .. (-1.090000,0.239400)
      .. controls (-1.105000,0.256083) and (-1.119792,0.271979) .. (-1.134375,0.287109)
      .. controls (-1.148958,0.302240) and (-1.163333,0.316604) .. (-1.177500,0.330225)
      .. controls (-1.191667,0.343846) and (-1.205625,0.356723) .. (-1.219375,0.368878)
      .. controls (-1.233125,0.381033) and (-1.246667,0.392467) .. (-1.260000,0.403200)
      .. controls (-1.273333,0.413933) and (-1.286458,0.423967) .. (-1.299375,0.433322)
      .. controls (-1.312292,0.442677) and (-1.325000,0.451354) .. (-1.337500,0.459375)
      .. controls (-1.350000,0.467396) and (-1.362292,0.474760) .. (-1.374375,0.481491)
      .. controls (-1.386458,0.488221) and (-1.398333,0.494317) .. (-1.410000,0.499800)
      .. controls (-1.421667,0.505283) and (-1.433125,0.510154) .. (-1.444375,0.514434)
      .. controls (-1.455625,0.518715) and (-1.466667,0.522404) .. (-1.477500,0.525525)
      .. controls (-1.488333,0.528646) and (-1.498958,0.531198) .. (-1.509375,0.533203)
      .. controls (-1.519792,0.535208) and (-1.530000,0.536667) .. (-1.540000,0.537600)
      .. controls (-1.550000,0.538533) and (-1.559792,0.538942) .. (-1.569375,0.538847)
      .. controls (-1.578958,0.538752) and (-1.588333,0.538154) .. (-1.597500,0.537075)
      .. controls (-1.606667,0.535996) and (-1.615625,0.534435) .. (-1.624375,0.532416)
      .. controls (-1.633125,0.530396) and (-1.641667,0.527917) .. (-1.650000,0.525000)
      .. controls (-1.658333,0.522083) and (-1.666458,0.518729) .. (-1.674375,0.514959)
      .. controls (-1.682292,0.511190) and (-1.690000,0.507004) .. (-1.697500,0.502425)
      .. controls (-1.705000,0.497846) and (-1.712292,0.492873) .. (-1.719375,0.487528)
      .. controls (-1.726458,0.482183) and (-1.733333,0.476467) .. (-1.740000,0.470400)
      .. controls (-1.746667,0.464333) and (-1.753125,0.457917) .. (-1.759375,0.451172)
      .. controls (-1.765625,0.444427) and (-1.771667,0.437354) .. (-1.777500,0.429975)
      .. controls (-1.783333,0.422596) and (-1.788958,0.414910) .. (-1.794375,0.406941)
      .. controls (-1.799792,0.398971) and (-1.805000,0.390717) .. (-1.810000,0.382200)
      .. controls (-1.815000,0.373683) and (-1.819792,0.364904) .. (-1.824375,0.355884)
      .. controls (-1.828958,0.346865) and (-1.833333,0.337604) .. (-1.837500,0.328125)
      .. controls (-1.841667,0.318646) and (-1.845625,0.308948) .. (-1.849375,0.299053)
      .. controls (-1.853125,0.289158) and (-1.856667,0.279067) .. (-1.860000,0.268800)
      .. controls (-1.863333,0.258533) and (-1.866458,0.248092) .. (-1.869375,0.237497)
      .. controls (-1.872292,0.226902) and (-1.875000,0.216154) .. (-1.877500,0.205275)
      .. controls (-1.880000,0.194396) and (-1.882292,0.183385) .. (-1.884375,0.172266)
      .. controls (-1.886458,0.161146) and (-1.888333,0.149917) .. (-1.890000,0.138600)
      .. controls (-1.891667,0.127283) and (-1.893125,0.115879) .. (-1.894375,0.104409)
      .. controls (-1.895625,0.092940) and (-1.896667,0.081404) .. (-1.897500,0.069825)
      .. controls (-1.898333,0.058246) and (-1.898958,0.046623) .. (-1.899375,0.034978)
      .. controls (-1.899792,0.023333) and (-1.900000,0.011667) .. (-1.900000,0.000000)
      .. controls (-1.900000,-0.011667) and (-1.899792,-0.023333) .. (-1.899375,-0.034978)
      .. controls (-1.898958,-0.046623) and (-1.898333,-0.058246) .. (-1.897500,-0.069825)
      .. controls (-1.896667,-0.081404) and (-1.895625,-0.092940) .. (-1.894375,-0.104409)
      .. controls (-1.893125,-0.115879) and (-1.891667,-0.127283) .. (-1.890000,-0.138600)
      .. controls (-1.888333,-0.149917) and (-1.886458,-0.161146) .. (-1.884375,-0.172266)
      .. controls (-1.882292,-0.183385) and (-1.880000,-0.194396) .. (-1.877500,-0.205275)
      .. controls (-1.875000,-0.216154) and (-1.872292,-0.226902) .. (-1.869375,-0.237497)
      .. controls (-1.866458,-0.248092) and (-1.863333,-0.258533) .. (-1.860000,-0.268800)
      .. controls (-1.856667,-0.279067) and (-1.853125,-0.289158) .. (-1.849375,-0.299053)
      .. controls (-1.845625,-0.308948) and (-1.841667,-0.318646) .. (-1.837500,-0.328125)
      .. controls (-1.833333,-0.337604) and (-1.828958,-0.346865) .. (-1.824375,-0.355884)
      .. controls (-1.819792,-0.364904) and (-1.815000,-0.373683) .. (-1.810000,-0.382200)
      .. controls (-1.805000,-0.390717) and (-1.799792,-0.398971) .. (-1.794375,-0.406941)
      .. controls (-1.788958,-0.414910) and (-1.783333,-0.422596) .. (-1.777500,-0.429975)
      .. controls (-1.771667,-0.437354) and (-1.765625,-0.444427) .. (-1.759375,-0.451172)
      .. controls (-1.753125,-0.457917) and (-1.746667,-0.464333) .. (-1.740000,-0.470400)
      .. controls (-1.733333,-0.476467) and (-1.726458,-0.482183) .. (-1.719375,-0.487528)
      .. controls (-1.712292,-0.492873) and (-1.705000,-0.497846) .. (-1.697500,-0.502425)
      .. controls (-1.690000,-0.507004) and (-1.682292,-0.511190) .. (-1.674375,-0.514959)
      .. controls (-1.666458,-0.518729) and (-1.658333,-0.522083) .. (-1.650000,-0.525000)
      .. controls (-1.641667,-0.527917) and (-1.633125,-0.530396) .. (-1.624375,-0.532416)
      .. controls (-1.615625,-0.534435) and (-1.606667,-0.535996) .. (-1.597500,-0.537075)
      .. controls (-1.588333,-0.538154) and (-1.578958,-0.538752) .. (-1.569375,-0.538847)
      .. controls (-1.559792,-0.538942) and (-1.550000,-0.538533) .. (-1.540000,-0.537600)
      .. controls (-1.530000,-0.536667) and (-1.519792,-0.535208) .. (-1.509375,-0.533203)
      .. controls (-1.498958,-0.531198) and (-1.488333,-0.528646) .. (-1.477500,-0.525525)
      .. controls (-1.466667,-0.522404) and (-1.455625,-0.518715) .. (-1.444375,-0.514434)
      .. controls (-1.433125,-0.510154) and (-1.421667,-0.505283) .. (-1.410000,-0.499800)
      .. controls (-1.398333,-0.494317) and (-1.386458,-0.488221) .. (-1.374375,-0.481491)
      .. controls (-1.362292,-0.474760) and (-1.350000,-0.467396) .. (-1.337500,-0.459375)
      .. controls (-1.325000,-0.451354) and (-1.312292,-0.442677) .. (-1.299375,-0.433322)
      .. controls (-1.286458,-0.423967) and (-1.273333,-0.413933) .. (-1.260000,-0.403200)
      .. controls (-1.246667,-0.392467) and (-1.233125,-0.381033) .. (-1.219375,-0.368878)
      .. controls (-1.205625,-0.356723) and (-1.191667,-0.343846) .. (-1.177500,-0.330225)
      .. controls (-1.163333,-0.316604) and (-1.148958,-0.302240) .. (-1.134375,-0.287109)
      .. controls (-1.119792,-0.271979) and (-1.105000,-0.256083) .. (-1.090000,-0.239400)
      .. controls (-1.075000,-0.222717) and (-1.059792,-0.205246) .. (-1.044375,-0.186966)
      .. controls (-1.028958,-0.168685) and (-1.013333,-0.149596) .. (-0.997500,-0.129675)
      .. controls (-0.981667,-0.109754) and (-0.965625,-0.089002) .. (-0.949375,-0.067397)
      .. controls (-0.933125,-0.045792) and (-0.916667,-0.023333) .. (-0.900000,0.000000)
      .. controls (-0.883333,0.023333) and (-0.866458,0.047542) .. (-0.849375,0.072647)
      .. controls (-0.832292,0.097752) and (-0.815000,0.123754) .. (-0.797500,0.150675)
      .. controls (-0.780000,0.177596) and (-0.762292,0.205435) .. (-0.744375,0.234216)
      .. controls (-0.726458,0.262996) and (-0.708333,0.292717) .. (-0.690000,0.323400)
      .. controls (-0.671667,0.354083) and (-0.653125,0.385729) .. (-0.634375,0.418359)
      .. controls (-0.615625,0.450990) and (-0.596667,0.484604) .. (-0.577500,0.519225)
      .. controls (-0.558333,0.553846) and (-0.538958,0.589473) .. (-0.519375,0.626128)
      .. controls (-0.499792,0.662783) and (-0.480000,0.700467) .. (-0.460000,0.739200)
      .. controls (-0.440000,0.777933) and (-0.419792,0.817717) .. (-0.399375,0.858572)
      .. controls (-0.378958,0.899427) and (-0.358333,0.941354) .. (-0.337500,0.984375)
      .. controls (-0.316667,1.027396) and (-0.295625,1.071510) .. (-0.274375,1.116741)
      .. controls (-0.253125,1.161971) and (-0.231667,1.208317) .. (-0.210000,1.255800)
      .. controls (-0.093000,1.512210) and (0.000000,1.750000) .. (0.000000,2.000000);
}
\newcommand{\DrawRibbonCenterOverpass}[1]{%
    \draw[#1] (-1.777500,-0.429975)
      .. controls (-1.771667,-0.437354) and (-1.765625,-0.444427) .. (-1.759375,-0.451172)
      .. controls (-1.753125,-0.457917) and (-1.746667,-0.464333) .. (-1.740000,-0.470400)
      .. controls (-1.733333,-0.476467) and (-1.726458,-0.482183) .. (-1.719375,-0.487528)
      .. controls (-1.712292,-0.492873) and (-1.705000,-0.497846) .. (-1.697500,-0.502425)
      .. controls (-1.690000,-0.507004) and (-1.682292,-0.511190) .. (-1.674375,-0.514959)
      .. controls (-1.666458,-0.518729) and (-1.658333,-0.522083) .. (-1.650000,-0.525000)
      .. controls (-1.641667,-0.527917) and (-1.633125,-0.530396) .. (-1.624375,-0.532416)
      .. controls (-1.615625,-0.534435) and (-1.606667,-0.535996) .. (-1.597500,-0.537075)
      .. controls (-1.588333,-0.538154) and (-1.578958,-0.538752) .. (-1.569375,-0.538847)
      .. controls (-1.559792,-0.538942) and (-1.550000,-0.538533) .. (-1.540000,-0.537600)
      .. controls (-1.530000,-0.536667) and (-1.519792,-0.535208) .. (-1.509375,-0.533203)
      .. controls (-1.498958,-0.531198) and (-1.488333,-0.528646) .. (-1.477500,-0.525525)
      .. controls (-1.466667,-0.522404) and (-1.455625,-0.518715) .. (-1.444375,-0.514434)
      .. controls (-1.433125,-0.510154) and (-1.421667,-0.505283) .. (-1.410000,-0.499800)
      .. controls (-1.398333,-0.494317) and (-1.386458,-0.488221) .. (-1.374375,-0.481491)
      .. controls (-1.362292,-0.474760) and (-1.350000,-0.467396) .. (-1.337500,-0.459375)
      .. controls (-1.325000,-0.451354) and (-1.312292,-0.442677) .. (-1.299375,-0.433322)
      .. controls (-1.286458,-0.423967) and (-1.273333,-0.413933) .. (-1.260000,-0.403200)
      .. controls (-1.246667,-0.392467) and (-1.233125,-0.381033) .. (-1.219375,-0.368878)
      .. controls (-1.205625,-0.356723) and (-1.191667,-0.343846) .. (-1.177500,-0.330225)
      .. controls (-1.163333,-0.316604) and (-1.148958,-0.302240) .. (-1.134375,-0.287109)
      .. controls (-1.119792,-0.271979) and (-1.105000,-0.256083) .. (-1.090000,-0.239400)
      .. controls (-1.075000,-0.222717) and (-1.059792,-0.205246) .. (-1.044375,-0.186966)
      .. controls (-1.028958,-0.168685) and (-1.013333,-0.149596) .. (-0.997500,-0.129675)
      .. controls (-0.981667,-0.109754) and (-0.965625,-0.089002) .. (-0.949375,-0.067397)
      .. controls (-0.933125,-0.045792) and (-0.916667,-0.023333) .. (-0.900000,0.000000)
      .. controls (-0.883333,0.023333) and (-0.866458,0.047542) .. (-0.849375,0.072647)
      .. controls (-0.832292,0.097752) and (-0.815000,0.123754) .. (-0.797500,0.150675)
      .. controls (-0.780000,0.177596) and (-0.762292,0.205435) .. (-0.744375,0.234216)
      .. controls (-0.726458,0.262996) and (-0.708333,0.292717) .. (-0.690000,0.323400)
      .. controls (-0.671667,0.354083) and (-0.653125,0.385729) .. (-0.634375,0.418359)
      .. controls (-0.615625,0.450990) and (-0.596667,0.484604) .. (-0.577500,0.519225)
      .. controls (-0.558333,0.553846) and (-0.538958,0.589473) .. (-0.519375,0.626128)
      .. controls (-0.499792,0.662783) and (-0.480000,0.700467) .. (-0.460000,0.739200)
      .. controls (-0.440000,0.777933) and (-0.419792,0.817717) .. (-0.399375,0.858572)
      .. controls (-0.378958,0.899427) and (-0.358333,0.941354) .. (-0.337500,0.984375)
      .. controls (-0.316667,1.027396) and (-0.295625,1.071510) .. (-0.274375,1.116741)
      .. controls (-0.253125,1.161971) and (-0.231667,1.208317) .. (-0.210000,1.255800)
;
}

%% file: figures/tikz-diagrams/s-matrix-diagrams.tex
\input{figures/tikz-diagrams/diagram-style.tex}
\usetikzlibrary{calc,decorations.markings,positioning}

\newcommand{\HopfCenteredArrow}[1][black]{%
  \fill[#1]
    (-1.75pt,-1.25pt) -- (2.05pt,0pt) -- (-1.75pt,1.25pt) -- cycle;%
}
\newcommand{\HopfReversedArrow}[1][black]{%
  \fill[#1]
    (1.75pt,-1.25pt) -- (-2.05pt,0pt) -- (1.75pt,1.25pt) -- cycle;%
}

\tikzset{
  appendix hopf left oriented/.style={
    appendix strand,
    postaction={decorate},
    decoration={
      markings,
      mark=at position 0.30 with {\HopfReversedArrow},
      mark=at position 0.70 with {\HopfReversedArrow}
    }
  },
  appendix hopf right oriented/.style={
    appendix strand,
    postaction={decorate},
    decoration={
      markings,
      mark=at position 0.20 with {\HopfCenteredArrow},
      mark=at position 0.80 with {\HopfCenteredArrow}
    }
  }
}

\newcommand{\HopfLinkAmplitude}[3][black]{%
  \begin{tikzpicture}[baseline=-0.55ex, x=0.82cm, y=0.82cm]
    \path[use as bounding box] (-1.52,-0.86) rectangle (1.52,0.86);
    \coordinate (HopfLeftCenter) at (-0.36,0);
    \coordinate (HopfRightCenter) at (0.36,0);
    \coordinate (HopfLeftLabel) at (-0.98,0);
    \coordinate (HopfRightLabel) at (0.98,0);

    \draw[appendix strand,draw=#1,postaction={decorate},
      decoration={markings,
        mark=at position .30 with {\HopfReversedArrow[#1]},
        mark=at position .70 with {\HopfReversedArrow[#1]}}]
      (HopfLeftCenter) circle[radius=0.62];
    \draw[appendix hopf right oriented]
      (HopfRightCenter) circle[radius=0.62];
    \draw[appendix crossing,draw=#1]
      ($(HopfLeftCenter)+(44:0.62)$)
      arc[start angle=44,end angle=65,radius=0.62];
    \draw[appendix crossing]
      ($(HopfRightCenter)+(224:0.62)$)
      arc[start angle=224,end angle=245,radius=0.62];

    \node[appendix label, left=2pt of HopfLeftLabel,text=#1] {$#2$};
    \node[appendix label, right=2pt of HopfRightLabel] {$#3$};
  \end{tikzpicture}%
}

